\documentclass[fleqn,usenatbib]{mnras}

\usepackage{newtxtext,newtxmath}

\usepackage[T1]{fontenc}
\usepackage{ae,aecompl}

\usepackage{graphicx}	
\usepackage{amsmath}	
\usepackage{amssymb}	
\usepackage{aecompl}
\usepackage[T1]{fontenc}

\usepackage{color}

\def\arcsec{$^{\prime\prime}$}

\def\lir{$L_{\rm IR}$}
\def\luv{$L_{\rm UV}$}

\def\Msun{$M_{\odot}$}

\def\Mstar{$M_{\ast}$}

\def\angstrom{\r{A}}
\def\uvbeta{$\beta_{\rm UV}$}

\title[Dust Attenuation of Massive Star-Forming Galaxies at $z\sim3-6$]
{The Dust Attenuation of Star-forming Galaxies at $z\sim3$ and Beyond: New Insights from ALMA Observations
}

\author[Fudamoto et al.]{
Y.~Fudamoto$^{1}$\thanks{E-mail: yoshinobu.fudamoto@unige.ch},
P. A.~Oesch$^{1}$,
E. Schinnerer$^{2}$,
B. Groves$^{3}$,
A. Karim$^{4}$,
\newauthor
B. Magnelli$^{4}$,
M. T. Sargent$^{5}$,
P. Cassata$^{6}$,
P. Lang$^{2}$,
D. Liu$^{2}$,
O. Le F\`{e}vre$^{7}$,
\newauthor
S. Leslie$^{2}$,
V. Smol\v{c}i\'{c}$^{8}$,
L. Tasca$^{7}$
\\
$^{1}$Observatoire de Gen\`{e}ve, 51 Ch. des Maillettes, 1290 Versoix, Switzerland\\
$^{2}$Max Planck Institute for Astronomy, K\"{o}nigstuhl 17, 69117 Heidelberg, Germany\\
$^{3}$Research School of Astronomy and Astrophysics, Australian National University, Canberra, ACT 2611, Australia\\
$^{4}$Argelander-Institut f\"ur Astronomie, Universit\"at Bonn, Auf dem H\"ugel 71, D-53121 Bonn, Germany\\
$^{5}$Astronomy Centre, Department of Physics and Astronomy, University of Sussex, Brighton, BN1 9QH, UK\\
$^{6}$Instituto de Fisica y Astronoma, Facultad de Ciencias, Universidad de Valparaiso, 1111 Gran Bretana, Playa Ancha Valparaso, Chile\\
$^{7}$Aix-Marseille Universit\'{e}, CNRS, LAM (Laboratoire d'Astrophysique de Marseille) UMR 7326, 13388 Marseille, France\\
$^{8}$Faculty of Science University of Zagreb Bijeni\v{c}ka c. 32, 10002 Zagreb, Croatia
}

\date{Accepted XXX. Received YYY; in original form ZZZ}

\pubyear{2017}

\begin{document}
\label{firstpage}
\pagerange{\pageref{firstpage}--\pageref{lastpage}}
\maketitle

\begin{abstract}
We present results on the dust attenuation of galaxies at redshift $\sim3-6$ by studying the relationship between the UV spectral slope ($\beta_{\rm UV}$) and the infrared excess (IRX; \lir/\luv) using ALMA far-infrared continuum observations. Our study is based on a sample of 67 massive, star-forming galaxies with a median mass of \Mstar$\sim 10^{10.7}\,$\Msun\ spanning a redshift range $z=2.6-3.7$ (median $z=3.2$) that were observed with ALMA at $\lambda_{rest}=300\,{\rm \mu m}$. Both the individual ALMA detections (41 sources) and stacks including all galaxies show the IRX--\uvbeta\ relationship at $z\sim3$ is mostly consistent with that of local starburst galaxies on average. However, we find evidence for a large dispersion around the mean relationship by up to $\pm0.5$ dex. Nevertheless, the locally calibrated dust correction factors based on the IRX--\uvbeta\ relation are on average applicable to main-sequence $z\sim3$ galaxies. This does not appear to be the case at even higher redshifts, however. Using public ALMA observations of $z\sim4-6$ galaxies we find evidence for a significant evolution in the IRX--\uvbeta\ and the IRX--\Mstar~relations beyond $z\sim3$ toward lower IRX values. We discuss several caveats that could affect these results, including the assumed dust temperature. ALMA observations of larger $z>3$ galaxy samples will be required to confirm this intriguing redshift evolution.
\end{abstract}




\section{Introduction}
One of the keys to understand the formation and evolution of galaxies is to estimate
 the star-formation activity of galaxies across cosmic times.
 Recent optical/near-infrared (NIR) observations successfully measured the star formation rate density (SFRD)
 of galaxies out to $z\sim10$ \citep[e.g.][]{Oesch2013,Oesch2014,Ellis2013,Bouwens2015,McLeod2016}, and showed that the SFRD steadily increases toward lower redshift, peaks around $z\sim2$--3, and
 then declines by an order of magnitude over the last seven billion years \citep[e.g.][]{Hopkins2006,Madau2014}.

However, a major caveat of current studies at $z>3$ in particular is that most measurements of the SFRD rely on the rest-frame UV light from young, massive stars to estimate the star formation rate
of galaxies. This UV light is highly sensitive to obscuration by dust particles in the interstellar medium. 
The absorbed energy in the UV is re-emitted in infrared (IR) wavebands by the heated dust. As a result, a galaxy emits a considerable amount of energy in the IR, which together with the UV luminosity provides a good measurement of the total star-formation of a galaxy \citep[e.g.][]{Elbaz2011,Symeonidis2013,Ivison2016}.
Without a direct detection of dust emission, however, one requires a well calibrated relation between 
total energy output from star formation activity and observed energy in UV
to recover the total star formation rate (SFR) of galaxies.

 The useful quantities that parametrise the amount of obscuration are the UV spectral slope (\uvbeta) 
 and the infrared excess (${\rm IRX}\equiv$\lir/\luv).
 \uvbeta~is defined as a power law form of  $f_{\lambda}\propto \lambda^{\beta_{\rm UV}}$ and thus describes galaxy colours in the UV waveband, which are modulated by dust attenuation. Thus one naturally expects a relation between the \uvbeta\ and
 the fraction of energy emitted by dust in the IR (i.e. ${\rm IRX}$).
 The IRX--\uvbeta~relation has been well calibrated using local star bursting galaxies
 \citep{Meurer1999,Overzier2011,Takeuchi2012,Casey2014}. In particular, the local relation estimated in \citet{Meurer1999} is very widely used, and we will refer to it as M99 in the following.

The locally calibrated IRX--\uvbeta~relation has been shown to be applicable to the general star-forming galaxy population at redshift $z=0$--2 \citep[e.g.][]{Reddy2006,Daddi2009,Pannella2009,Reddy2010,Overzier2011,Sklias2014}.
However, it is not clear if the relation is still valid at $z\gtrsim3$, as
the sample of individual IR detections is limited to the most luminous galaxies with extremely high SFR \citep[e.g.][]{Oteo2013,Riechers2013}.

Despite the power of stacking analyses of high-redshift galaxies
using single dish sub/mm telescopes \citep{Heinis2013,Heinis2014,Pannella2015,Alvarez-Marquez2016,Bourne2017}, detailed studies 
require individual detections of IR emission from normal star-forming galaxies along the star-forming main sequence.
 
 With currently accumulating observations with extremely sensitive sub/mm interferometers
  such as ALMA and NOEMA, the study of dust emission
  from high-redshift star forming galaxies is rapidly maturing.
 These observations hint at a potential evolution of the IRX--\uvbeta~relation at $z\gtrsim3$,
 with high-redshift star forming galaxies appearing to be 'infrared-dark' \citep{Ouchi2013,Ota2014,Capak2015,Schaerer2015,Bouwens2016,Dunlop2017}.
However, current studies are still based on small samples only, and the direct detection of IR emission from statistical samples of individual high-redshift galaxies is still missing.

In this paper, to improve our knowledge of the nature of dust emission from $z\gtrsim3$
 star-forming galaxies, we explore the IRX--\uvbeta~relation at $z\sim3.2$ by using our ALMA observations of 
UV selected, massive (\Mstar$\gtrsim10^{10.7}$\Msun), star-forming galaxies located on COSMOS field.
 We also include public ALMA observations of $z\sim4$--6 galaxies,
 in order to investigate a possible redshift evolution of the dust attenuation between $z\sim3$ and $z\sim6$, within the first 2 Gyr of cosmic history.
 
 This paper is organised as follows: In Section 2, we present our galaxy sample and the data used for the analysis. Section 3 describes the basic measurements performed, before our results are presented in Section 4 and we conclude in Section 5.
  Throughout this paper, we assume a cosmology  with $(\Omega_{m},\Omega_{\Lambda},h_0) = (0.3,0.7,0.7)$, and a Chabrier initial mass function \citep{Chabrier2003} where applicable.

\section{Observations}
 \subsection{ALMA Observations of $z\sim3.2$ Targets}
 The main data used for this study come from an ALMA Cycle-2 Program (2013.1.00151.S, PI E.~Schinnerer),
 which targets the $\sim\,300\,\rm{\mu m}$ rest-frame continuum of a sample of main-sequence galaxies at $z\sim3$. Details of the target selection and ALMA observations are presented in \citet{Schinnerer2016}, and they are briefly summarised below.

The main purpose of the ALMA program was to observe massive, star-forming galaxies at $z\sim3$ on the main sequence. 
In this paper, we only include sources with reliable photometric redshifts in the range 2.6--3.7
  in the latest {\sc COSMOS15} catalogue \citep{Laigle2016}
 and we exclude sources that potentially host an AGN \citep[as discussed in][]{Schinnerer2016}. Our final $z\sim3$ sample thus consists of 67 galaxies,
 which have a median redshift of 3.18,
 and a median stellar mass of ${\rm log(M_{\ast} [M_{\odot}]) \sim 10.7}$,
 with the minimum/maximum stellar mass being ${\rm log(M_{\ast} [M_{\odot}]) = 9.2/11.5}$.
 While most sources only have photometric redshifts, 7 galaxies of our sample have high-quality spectroscopic redshift measurements from VIMOS Ultra Deep Survey \citep[VUDS;][]{LeFevre2015}.
 
 ALMA observations of these targets were performed to detect the continuum emission at $240\,\rm{GHz}$,
 which corresponds to rest-frame $\lambda_{\rm rf}\sim\,300\,\rm{\mu m}$ at the redshift of our sample. 
 Targets were observed with typically 38 antennas between December, 25th and 30th, 2015 for an average on-source time of 2\,min. The achieved beam-size and rms of the ALMA images are $1.8$\arcsec$\times$1.1\arcsec(1.7\arcsec$\times$1.1\arcsec)
 and $66(77)\,\rm{\mu Jy/beam}$ for 19(48) 
 fields using the {\sc NATURAL} weighting scheme.

 A total of 41 of our 67 targets were detected at more than $>3\,\sigma$ at the counterpart positions of the latest {\sc COSMOS} Ultra-Vista NIR images \citep{Mccracken2012}. Four out of these sources have reliable spectroscopic redshifts from VUDS.
 For the 26 targets with ALMA non-detections, we compute $3\sigma$ upper limits on the ALMA fluxes that are used throughout the paper.

 Stellar masses for our $z\sim3$ sample are calculated based on multi-wavelength spectral energy distribution (SED) fitting using the {\sc MAGPHYS} code \citep{Dacunha2008}. In addition to the COSMOS optical+NIR photometry \citep{Laigle2016}, we also exploit all available longer wavelength data in these fits. This includes the Spitzer/MIPS $24\,{\rm \mu m}$ fluxes from \citet{LeFloch2009},
 the Herschel/PACS $100\, \mu m$ fluxes from \citet{Lutz2011}, and our ALMA 1.3-mm fluxes \citep[see][for details on the SED fitting]{Schinnerer2016}. 

 \subsection{$z\ge4$ Galaxy Samples}
 While our main analysis is based on the $z\sim3$ sample from \citet{Schinnerer2016},
 we also investigate a potential redshift evolution of the IRX--\uvbeta~relation in section \ref{redev}.
 To do this, we exploit the published ALMA fluxes of $z\sim4$--5 galaxies from \citet{Scoville2016},
 and the IRX--\uvbeta~data points of $z\sim5$--6 galaxies from \citet{Capak2015}.

ALMA band-6 fluxes of the $z\sim4-5$ sample are taken from Table 8 of \citet[][]{Scoville2016}. This sample consists of 19 galaxies with median $\log(M_{\ast} [M_{\odot}])$ of 10.45 and median redshift of 4.3.
For this sample, the observed ALMA fluxes probe rest-frame wavelengths $\lambda_{\rm rf}\sim240\,{\rm \mu m}$. The UV luminosities and continuum slopes are derived in the same manner as for our $z\sim3$ sample based on the photometry of the {\sc COSMOS15} catalogue. Stellar masses for the $z\sim4-5$ sources are directly taken from that catalogue.

For further constraints at even higher redshift, we exploit the data set by \citet[][]{Capak2015}.
That sample consists of 9 galaxies (4 detections and 5 non-detections) with median mass $\log(M_{\ast} [M_{\odot}])$ of 9.86 and median redshift of 5.55.
We utilise the \lir, \luv, and \uvbeta\ estimates of the targets from Table 1 of \citet{Barisic2017}, which updates the original measurements by using additional HST photometry (and also provides measurements for individual clumps in two sources).
We excluded the X-ray detected quasar from their sample.

 \begin{figure}
\includegraphics[width=\columnwidth]{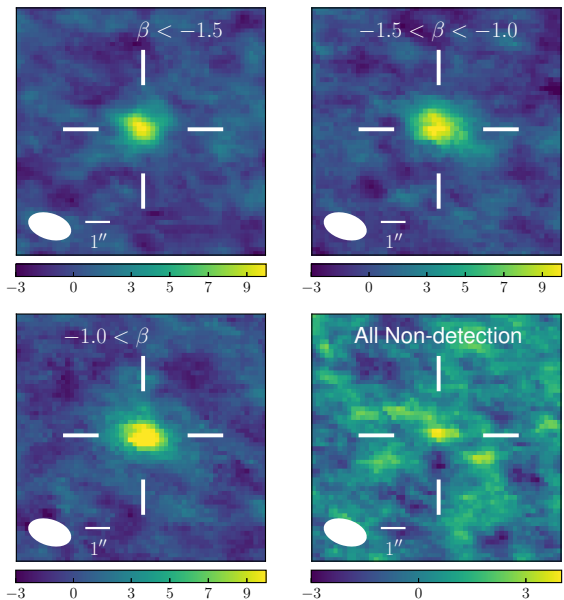}
\caption{Stacked images of our ALMA 1.3-mm continuum observations for different sub-samples.
	$1^{\prime\prime}$ scales and our synthesised beam FWHM sizes are shown at the bottom left corners of each panel.
	Median stacks are computed using all $z\sim3.2$ ALMA sources (individually detected and non-detected) split into 3 bins of UV slope (top left: $\beta_{\rm UV} < -1.5$ with 14 galaxies,
	top right: $-1.5<\beta_{\rm UV}<-1.0$ with 30 observations,
	and bottom left: $-1.0<\beta_{\rm UV}$ with 23 observations). All stacks result in a clear detection.
	The bottom right panel corresponds to a stack of all individual ALMA non-detections (26 sources), for which we find a tentative detection with $\sigma\sim4.3$.
	We note that this is the most significant positive/negative signal within the primary beam FWHM of
	the stack. Colour bars below each panel indicate the signal to noise ratio per pixel. 
	The stacking results are summarised in Tab. \ref{res-stack}.
	}
\label{stack}
\end{figure}

\section{Analysis}

The analysis of the IRX--\uvbeta~relation requires us to measure the galaxies' luminosities
at UV and IR wavelengths, as well as the UV continuum slope \uvbeta, which we describe in the following section.
  
 \subsection{\lir, \luv, and \uvbeta}
 Infrared luminosities (\lir) are estimated by fitting a single component modified black body function to the ALMA $240\,{\rm GHz}$ fluxes of our galaxies and integrating over the wavelength range
 between 8 and $1000\,{\rm \mu m}$.
 We assume a dust emissivity index of $\beta_{\rm{d}}=1.5$ \citep{Dunne2001}.

Dust temperatures ($T_d$) are estimated based on a stacking analysis of {\it Spitzer}, {\it Herschel}, and ALMA images of our full $z\sim3$ galaxy sample \citep[see][for a similar analysis]{Magnelli2014}.
From the full IR SED of these stacked data, we obtain a best-fit dust temperature of $T_d=41.5\pm2.5\,{\rm K}$ by fitting a standard modified black body spectrum \citep{Casey2012}.
The derived dust temperature is consistent with recently obtained values at $z\sim4$ galaxies \citep[e.g.][]{Schreiber2016}.
Hereafter, we will use $T_d=40\,{\rm K}$ as our fiducial value throughout the paper, but we will comment on the potential impact of assuming a lower dust temperature of $T_d=35$\,K.

 Uncertainties on our ${L_{\rm IR}}$ measurements are estimated using the ALMA flux measurement errors.
 For the 26 non-detected targets, we estimated upper limits on their \lir~ using the $3\sigma$ ALMA flux limits.
  
 The \lir~of our detected targets range between $5.9\times10^{11}$ and $5.8\times10^{12}\,L_{\odot}$, thus straddling the range between luminous infrared galaxies (LIRGs, $10^{11}<L_{\rm IR}<10^{12}\,L_{\odot}$) and
 ultra luminous infrared galaxies (ULIRGs, $10^{12}<L_{\rm IR}<10^{13}\,L_{\odot}$).

 While the increased CMB temperature can affect the luminosity measurements of higher redshift galaxies \citep[e.g.][]{Dacunha2013,zhang2016}, we estimate that the expected effect on our $z\sim3$ sample is still less than 5\% and thus negligible compared to our typical flux measurement uncertainties.

 UV spectral slopes ($\beta_{\rm{UV}}$) are calculated
 by fitting a power law $f_{\rm{\lambda}}\propto \lambda^{\beta_{\rm{UV}}}$ to 
 the broad- and narrow-band photometry of the galaxies over the rest-frame wavelength range $1500-2500\,$\angstrom.
 For this purpose, we utilise the photometry measured within 3\arcsec\ apertures listed in the {\sc COSMOS15} catalogue \citep{Laigle2016}.
 Typically more than 10 data points are available covering this waveband for each of our targets.
 The monochromatic UV luminosities are then calculated at rest-frame $1600\,$\angstrom~using the above power law fits to the UV SEDs and defining $L_{\rm{UV}} = \nu_{1600}\,L_{\nu_{1600}}$.
 The range of obtained \luv~covers $1.2\times10^{10}<\nu_{1600}\,L_{\nu_{1600}}<1.5\times10^{11}\,L_{\odot}$.

\begin{table}
	\begin{center}
	\caption{Results of the Stacking Analysis }
	\label{res-stack}
	\begin{tabular}{ccccc}
		\hline\hline
		\uvbeta & \# of sources & log \Mstar$^a$&log \lir$^b$& logIRX\\
		&&[$M_{\sun}$]&[$L_{\sun}$]&\\
		\hline\\
		\multicolumn{5}{c}{All sources}\\
		$<-1.5$&14&$10.7$&$11.8$&$0.97^{+0.12}_{-0.12}$\\
		$-1.5$ to $-1.0$&30&$10.6$&$11.9$&$1.11^{+0.07}_{-0.08}$\\
		$>-1.0$&23&$10.8$&$12.0$&$1.28^{+0.06}_{-0.09}$\\\\
		\multicolumn{5}{c}{Non-detection sources}\\
		All&26&$10.7$&$11.1$&$0.37^{+0.11}_{-0.09}$\\
		\hline
	\end{tabular}
	\end{center}
	$\rm^{a}$ Median stellar mass of the target galaxies as estimated by the {\sc MAGPHYS} code \citep{Dacunha2008}.\\
	$\rm^{b}$ Corresponding IR luminosity of the stacked ALMA images.
\end{table}

\subsection{Stacking Analysis}

While 41 out of our sample of 67 galaxies are detected with ALMA, the remaining targets still contain important information on the average IRX relation at $z\sim3$. We therefore perform a stacking analysis of all target fields. 
 In particular, we are interested in the average IRX of our sample as a function of UV continuum slope \uvbeta.
 We thus split our full galaxy sample in three bins: 
 \uvbeta~$<-1.5,~$\uvbeta$\,=-1.5$ to $-1.0$, and \uvbeta~$>-1.0$, which contain 14, 30, and 23 galaxies, respectively.
 
 Thanks to the very uniform synthesised beam full-width-at-half-maximum (FWHM) of our ALMA images, stacks can be created by simply taking the median of each pixel in the ALMA maps. 
 During the stacking procedure, the ALMA images are weighted
  by the inverse of the galaxy's \luv, in order to compute the median IRX of the sample going into the stack. All three ALMA stacks reveal a significant detection, as shown in Fig. \ref{stack}.

 We also compute a stacked image of all 26 ALMA non-detections
 using the same procedure as above.
 The stack reveals a tentative detection with a
 significance of $\sim4.3\,\sigma$ (lower right panel of Fig. \ref{stack}).
 We note that this tentative detection is the most significant positive/negative signal
 within the central $23^{\prime\prime}\times23^{\prime\prime}$ of the stacked image,
 corresponding to the FWHM of the primary beam of each observation at {240\,{GHz}}. 
 
  We then measure total fluxes for each of the stacks using the peak fluxes and applying a correction to total flux derived from the mean of the ALMA detected sources. 
  Uncertainties on all stacked quantities, including the IRX, are estimated using bootstrap resampling.
 The results of this stacking analysis are summarised in Table \ref{res-stack}, where we also tabulate the resulting median IRX of the different stacks that are used in our analysis later on.

\begin{figure}
\includegraphics[width=\columnwidth]{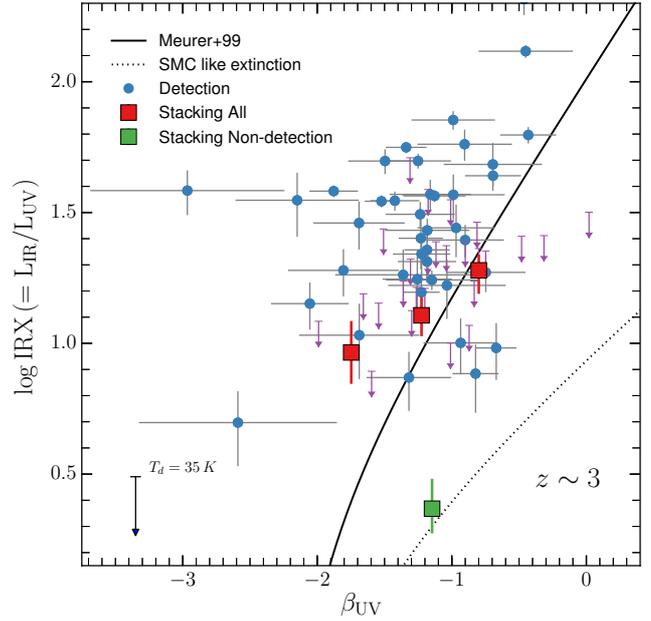}
\caption{The IRX--\uvbeta~diagram for massive (median \Mstar$\sim10^{10.7}$\Msun) star-forming galaxies at $z\sim3.2$
	using individual detections (blue dots), and $3\,\sigma$ upper limits for non-detections (magenta arrows). 
	Our stacking results are plotted with large squares, where the three red points correspond to median stacks of all galaxies in bins of \uvbeta\ representing the average IRX--\uvbeta\ relation of our sample. The green square corresponds to the median stack of all 26 ALMA non-detection fields.
	Two typically assumed attenuation curves are also shown: the relation of local starburst galaxies \citep[][solid line]{Meurer1999}, and an SMC-like relation \citep[e.g.][dotted line]{Prevot1984}.
	While our fiducial measurements assume a dust temperature ($T_d$) of $40\,{\rm K}$, the downward arrow in the lower left corner shows the impact of assuming $T_d=35\,{\rm K}$.
	While individual detections (and upper limits) typically lie above the M99 relation, our stacking analysis shows that the local M99 relation is generally applicable for the average $z\sim3$ galaxy. However, there is a considerable amount of dispersion around the mean IRX--\uvbeta~relation at $z\sim3.2$ with some sub-samples of galaxies (i.e. our individually non-detected sources) that are more consistent with an SMC-type attenuation curve.}
	
\label{ourirxbeta}
\end{figure}

\section{Results and Discussion}
In the following,  we discuss our results on the IRX--\uvbeta~and the 
IRX--\Mstar~relations at $z\sim3.2$, as well as their possible redshift evolution to $z\sim4-6$.

\subsection{IRX--\uvbeta~Relation at $z\sim3.2$}

The IRX--\uvbeta~diagram of our $z\sim3.2$ galaxy sample is shown in Fig. \ref{ourirxbeta} together with the expected relations of two different local dust extinction curves, the M99 relation and SMC-type extinction \citep[e.g.][]{Prevot1984}. These two attenuation laws have both been discussed in the past to represent different sub-samples of $z\sim2-3$ galaxies \citep[e.g.][]{Reddy2012,Bouwens2016}.

The figure shows that our ALMA detected targets typically lie above the M99 line. Similarly, the $3\sigma$ upper limits of individually un-detected sources often lie above the M99 relation.
However, when we combine these samples through stacks in bins of \uvbeta, we obtain three measurements of the IRX--\uvbeta\ relation that lie at the low end of the individually detected IRX values, showing that the typical $z\sim3$ galaxy is mostly consistent with the M99 relation, in particular at $\beta>-1.5$.

Since the sample with $\beta<-1.5$ only contains 14 sources, its stacked IRX value is quite uncertain. It lies significantly above the M99 relation. However, in the redder UV slope bins, where we have better statistics, both median stacked IRX values are in excellent agreement with the M99 relation.

Interestingly, the median stacks show that an SMC-like attenuation is clearly not applicable for the general galaxy sample at $z\sim3.2$. In the \uvbeta\ range probed by our stacks, the median IRX values lie 
$\gtrsim0.7\,{\rm dex}$ above an SMC-like attenuation curve.
However, considering the stack of ALMA non-detected sources only, we find a very low IRX value that lies within $<1\sigma$ of the SMC curve. 

This provides some evidence for a large dispersion in IRX values at a given \uvbeta\ from one galaxy to another at $z\sim3$, with certain subsamples of galaxies that show significantly lower IRX values at a given \uvbeta\ than the average population. This confirms previous analyses of $z\sim2-3$ galaxies which find that the local M99 relation is typically applicable to the general galaxy population, but that there are a few galaxies (that are typically younger) that are more consistent with an SMC-type attenuation \citep[][]{Siana2009,Reddy2012}. Within our sample, we do not find a clear separation in physical parameters, however, between our ALMA detections and non-detections (e.g. $z_{\rm phot}$, age, \Mstar). 

Note that our conclusions would not change significantly under the assumption of a lower dust temperature, $T_d = 35\,{\rm K}$ that has been used in some previous analyses of high-redshift galaxies \citep[e.g.,][]{Bouwens2016}. Under this assumption, the values of \lir~as well as the IRX decrease by $\sim0.25\,{\rm dex}$ for our sample (see arrow in Fig 2).

\begin{figure}
\includegraphics[width=\columnwidth]{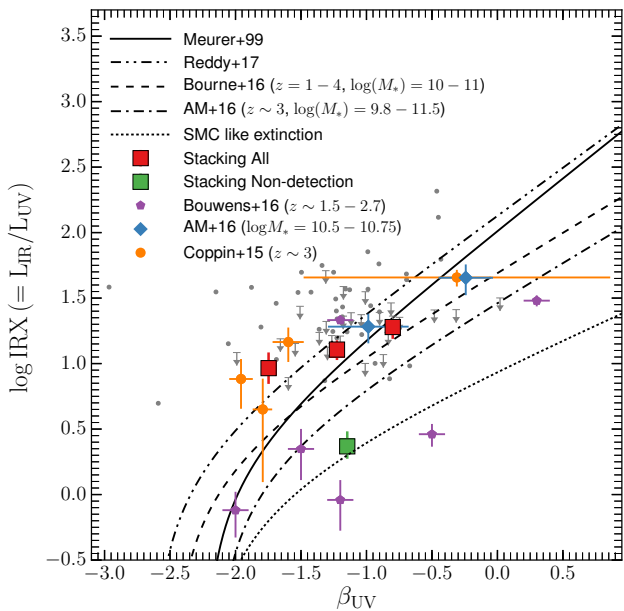}
\caption{IRX--$\beta_{\rm UV}$ diagram from several previous studies.
	Our results of individual detections and $3\,\sigma$ upper limits are 
	shown as grey dots and arrows. The results for stacked images are indicated by red squares
	(all sources) and a green square (for non-detected fields only).
	Results from previous studies include stacking analyses and 
	individual ALMA detections. {\it Blue}: stacking analysis of $z\sim3$ LBGs with mass range
	 ${\rm log(}M_{\ast})\sim10.5$--11.0 from \citet[][labelled AM+16]{Alvarez-Marquez2016}.
	 {\it Yellow}: Stacking results of LBGs at $z\sim3-4$ with \Mstar $\gtrsim10^{9.7}\,M_{\sun}$ \citep{Coppin2015}.
	{\it Purple}: individual detections of \Mstar$>10^{9.75}$\Msun\ galaxies at $z=2-3$ \citep{Bouwens2016}.
	Lines show different best-fit relations of local galaxies, as well as for a $z\sim3$ LBG sample ({\it dash-dotted} line; AM+16) and 
	IR selected $z\sim1-4$ galaxies \citep[dashed line;][the appropriate mass ranges are indicated in the label]{Bourne2017}.
	Finally, the {\it dash-double-dotted} line shows a theoretical relation based on an intrinsically very blue SED reddened by the \citet{Reddy2015} extinction curve that appears to provide the best representation of the IRX--$\beta$ of our full sample.
	}
\label{otherdata}
\end{figure}

\subsection{Comparison to Previous $z\sim3$ Studies}

In Fig. \ref{otherdata}, we compare our results with several previous studies of the
IRX--\uvbeta\ relation at $z\sim3$ for samples covering a similar mass range (\Mstar$\gtrsim10^{10}$\Msun). In particular, the comparison samples include stacking analyses of Lyman Break Galaxies (LBGs) and/or IR-selected galaxies \citep{Coppin2015,Bourne2017,Alvarez-Marquez2016,Reddy2017},
 as well as individual ALMA detections of $z\gtrsim1.5$--2.7 galaxies from \citet{Bouwens2016}.
 Despite similar mass ranges probed, these previous analyses found IRX values that are sometimes different by $\gtrsim1\,{\rm dex}$ at a fixed \uvbeta.
 
 Interestingly, our data points and stacks are in good agreement with the full range of these previous analyses.
For instance, the stack of our non-detections shows an IRX value
similar to individual galaxies found by \citet{Bouwens2016}.
On the other hand, our stacked measurements and our individual detections are in good agreement with the values found by \citet{Coppin2015},
who utilised a stacking analysis of SCUBA-2 and {\it Herschel}
observations of $\sim4200$ LBGs. Similar agreement is found with the $z\sim3$ high-mass stacks from \citet{Alvarez-Marquez2016}.

We also investigate whether an IRX--$\beta$ relation with bluer intrinsic UV slope could explain our data.
Such a trend to bluer intrinsic slopes at higher redshift is expected due to, e.g., lower metallicities and younger stellar population ages \citep[e.g.][]{Dayal2012,Schaerer2013,Alavi2014,Castellano2014,debarros2014,Sklias2014,Cullen2017}, and evidence for such bluer intrinsic slopes is continuously emerging from several studies. 
For instance, \citet{Smit2016} derive a similar best fit $\beta_{\rm UV, int}=-2.5$ by comparing UV-based SFRs with those inferred from ${\rm H_{\alpha}}$ fluxes at $z\sim4$.
\citet{Reddy2017} provide theoretical derivations of IRX--$\beta$ relationships based on different assumptions of the intrinsic SED and dust curves. A relation based on a young SED with $\beta_{\rm UV, int}=-2.62$, reddened by the \citet{Reddy2015} dust curve indeed provides an adequate representation of our stacked points, including the bluest bin, in better agreement than the M99 relation. It will thus be interesting to further investigate the overall shape of the IRX--\uvbeta~relation and the $\beta_{\rm UV,\,int}$ based on larger ALMA surveys of high redshift galaxies spanning a wide range in UV continuum slopes in the future. 

\begin{figure*}
\centerline{\includegraphics[width=6.8in]{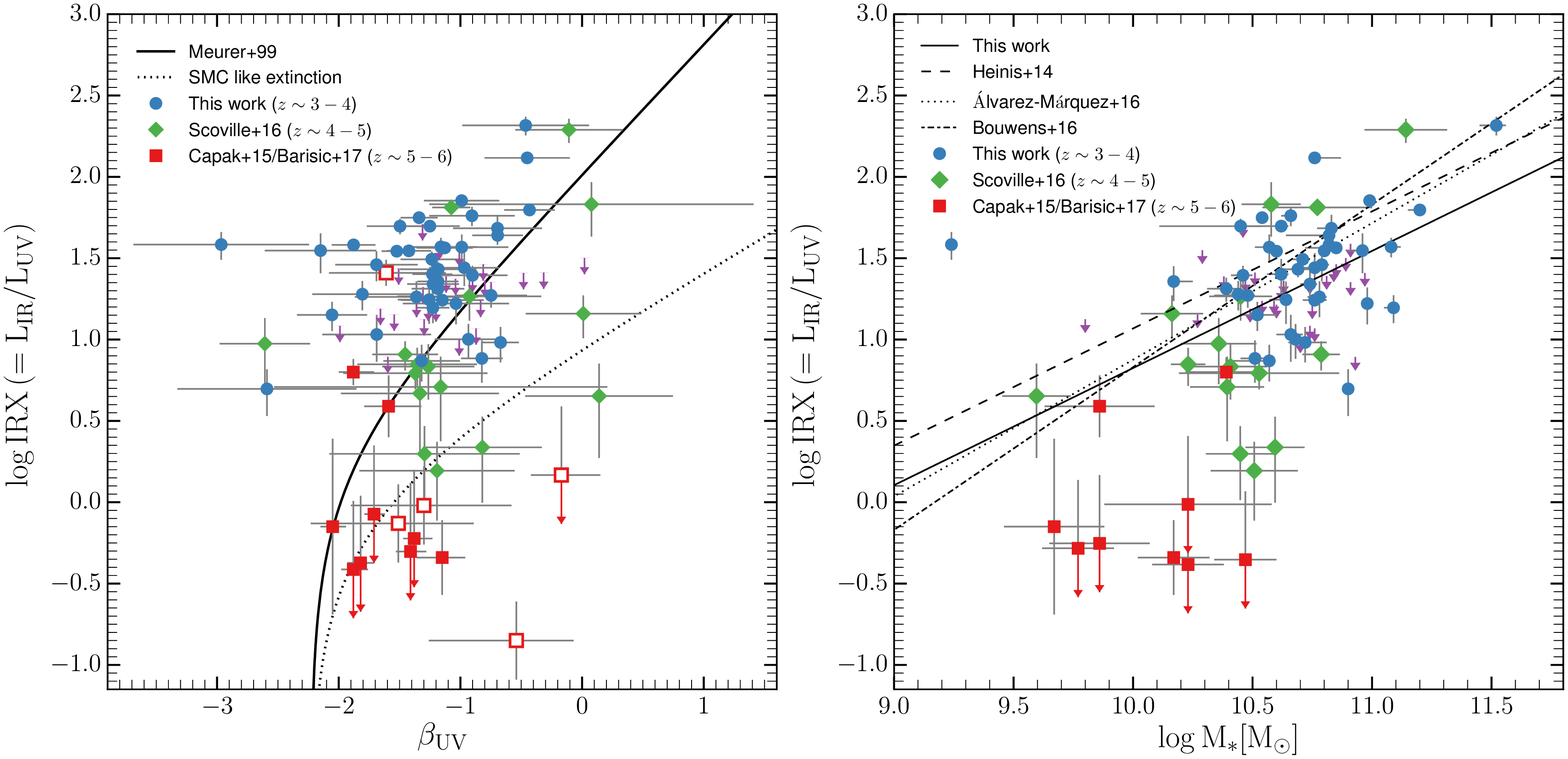}}
\caption{IRX-\uvbeta~and IRX--\Mstar~diagrams at $z=3-6$ utilising ancillary datasets from \citet{Scoville2016,Capak2015,Barisic2017}.
	Individual detections and $3\,\sigma$ upper limits are shown at
	at $z\sim3$--4 (blue circles and magenta arrows; this work), $z\sim4$--5 \citep[green diamonds, ][]{Scoville2016}, and 
	$z\sim5$--6 \citep[red squares, from][]{Capak2015,Barisic2017}.
	{\bf Left panel}: IRX-\uvbeta~diagram combining data at $z\sim3$--6.
	Lines show the IRX-\uvbeta~relation of local starburst galaxies \citep[solid, ][]{Meurer1999},
	and an SMC like extinction curve (dotted).
	For $z\sim5-6$ samples, we plot all data points including the measurements of individual clumps (empty red squares) as well as integrated values \citep[filled red squares; see][]{Barisic2017}.
	{\bf Right panel}: IRX-\Mstar~diagram of the same samples. Lines show stacking analyses of LBGs
	 at $z\sim3$ \citep[dotted line;][]{Alvarez-Marquez2016}, UV selected galaxies
	 at $z\sim 3$ \citep[dashed line;][]{Heinis2014}, $z\sim2-3$ galaxies in the HUDF \citep[dot-dashed line;][]{Bouwens2016}, and best-fit results to our data at $z\sim3.2$ (solid line).
	These figures suggest the existence of a significant redshift evolution of the IRX-\uvbeta~relation 
	between $z\sim3$ and $z\sim6$ even when the IRX--\Mstar~correlation is taken into account.
	}
\label{zev}	
\end{figure*}

\subsection{The IRX-\Mstar~Relation}
\label{irxmass}
Several studies have shown that more massive star-forming galaxies show larger dust attenuation than their lower mass counterparts, which results in a mass dependence of the IRX values \citep[e.g.][]{Reddy2010,Buat2012,Heinis2013,Coppin2015,Pannella2015,Bouwens2016,Bourne2017,Dunlop2017}.

In a recent analysis, \citet{Alvarez-Marquez2016} discuss the IRX--\Mstar~relation based on a stacking analysis of $z\sim3$ LBGs using {\it Herschel} and AzTEC data.
Although their mean IRX--\uvbeta~relation exhibits  a lower IRX than M99 at fixed \uvbeta~as shown in our Fig. \ref{otherdata},
their stacking analysis of a mass-matched sub-sample with \Mstar$\sim10^{10.5}$--$10^{11}$
results in a $\sim 0.5$ dex higher IRX at fixed \uvbeta~than
 their mean relation, agreeing with the M99 relation and our results (see Fig. \ref{otherdata}), and it clearly demonstrates the importance of comparing mass-matched galaxy samples.

In Fig. 4 right we analyse the IRX--\Mstar\ relation of our sample. Although the dynamic range in mass is small, there is generally a correlation with a slope that is consistent with previous studies. 
However, our targets show lower IRX values than previous studies at fixed \Mstar. We note that this could potentially be due to different \uvbeta\ distributions, in particular the lack of redder objects in our sample (i.e., with \uvbeta~$>-0.5$).

We quantify the offset by fitting for the normalisation but keeping the slope fixed to the one found in \citet{Heinis2013}. This results in a best-fit IRX--\Mstar\ relation following:
\[
     {\rm log(IRX)}=0.72\,[{\rm log(M_{\ast})}-10.35]+1.08
\]
The offset in normalisation relative to \citet{Heinis2013} is thus $0.24\,{\rm dex}$. A similar offset is found relative to the IRX--\Mstar\ relation from \citet[][]{Alvarez-Marquez2016}.

Interestingly, the sub-sample of our ALMA non-detections shows an even larger offset to lower IRX values compared to previous studies when we consider the location of their median stacked IRX.
This suggests that there is also a significant scatter
around the mean IRX--\Mstar~relation at $z\gtrsim3$.
In \citet{Reddy2010}, the authors discussed the IRX--\Mstar~relation of $z\sim2$ UV selected star forming galaxies, 
and estimated a scatter in the relation of $\sim0.46\,{\rm dex}$ about a linear fit.
Consistent with this, our result suggests that individual galaxies 
indeed can span a $\sim0.5\,{\rm dex}$ range in IRX values at $z\sim3$ around the IRX--$\beta_{\rm UV}$ relation. Such a large dispersion will result in large uncertainties on the inferred dust corrections of UV-based SFRs of individual galaxies, and it will be crucial to search for correlations with other physical parameters in an attempt to decrease these dispersions in the future with larger samples.
For instance, \citet{Nordon2013} show that the scatter in the IRX is correlated with the position of a galaxy with respect to the main-sequence of star-formation.

\subsection{The Evolution of IRX--\uvbeta~Beyond $z=3$}
\label{redev}
Our finding that the IRX--\uvbeta~relation at $z\sim3.2$ is consistent with the local M99 relation
is in contrast to some recent results on high redshift galaxies.

For instance, \citet{Capak2015} find that their sample of $z\sim5.5$ galaxies generally lies
$\sim1\,{\rm dex}$ below the M99 relation \citep[see also][]{Barisic2017}.
Similarly, \citet{Bouwens2016} find that the $z>4$ IRX--\uvbeta~relation lies below the M99 relation
based on a stacking analysis of Lyman Break Galaxies (LBGs) in the ASPECS data over the Hubble Ultra-Deep Field.

To investigate this evolution further, we compare our $z\sim3.2$ galaxies with the $z\sim4-5$ sample from \citet{Scoville2016}
and with the $z\sim5.5$ galaxies from \citet{Capak2015,Barisic2017} in Fig. \ref{zev}.
While the $z>4$ sample is currently very small, we see tentative evidence for an evolutionary trend at $z>3$.
Between $z\sim3$--6, the median IRX gradually drops by $\gtrsim1\,{\rm dex}$, and a large fraction of the $z\ge4$ sample lies close to or even below the SMC like extinction curve. 

This evolution to lower IRX values at a given \uvbeta\ is very intriguing, and could potentially hint at an evolution of the dust properties within the first 2 Gyr of cosmic history. This is expected theoretically due to the younger stellar population ages of galaxies in the early universe for which the AGB star population has not yet had enough time to produce a significant amount of dust. At very high redshift, the main dust production mechanism is thus thought to be supernovae explosions, which can be expected to lead to a different dust composition and attenuation curve than at later times \citep[][]{Todini2001,Gallerani2010}. Alternative solutions as to why higher redshift galaxies show lower dust emission could involve a higher fraction of gas to be locked up in a cold, molecular component compared to lower redshift \citep{Ferrara2016}. Furthermore, it will be important to understand the relative morphological distributions and the mixing of the stars and the dust in multi-component high redshift \citep[e.g.][]{Casey2014,Koprowski2016}.

However, before making strong claims about a possible redshift evolution, we have to control for the differences in stellar masses of the different samples. To do this, we explore the IRX--\Mstar~relation of the three high-redshift samples in the right panel of Fig. \ref{zev}.
As is evident from the plot, there is indeed a difference in mass among the different samples. However, the higher redshift samples generally show a larger offset from the $z\sim3$ stacking analysis of \citet[][]{Heinis2014} at lower masses.
The median offsets from our best-fit IRX--\Mstar\ relation are $\sim0.31\,{\rm dex}$ and $\sim1.2\,{\rm dex}$ for the
 samples at $z\sim4.3$ and $z\sim5.5$ (including $3\,\sigma$ upper limits), respectively.

Note that this rapid redshift evolution is in general agreement with the best-fit IRX--\Mstar~relation presented in \citet{Bouwens2016}, where the authors studied star forming galaxies with ${\rm log(}M_{\ast})<9.5$ at $z\sim4$--10, and found lower IRX values by $\sim0.5\,{\rm dex}$ at fixed \Mstar\ compared to a consensus relation estimated from previous analyses at $z\sim2-3$ \citep{Reddy2010,Whitaker2014,Alvarez-Marquez2016}. However, this was based on the assumption of $T_d=35\,{\rm K}$. When assuming that the dust temperature increases with redshift, their results are in better agreement with the previous relations (Fig \ref{zev}).

It is thus clear that it is critical to constrain the dust temperatures at $z>4$. Changes in $T_d$ can have a significant impact on the \lir\ and IRX values estimated from single-band observations,
and could affect all samples included in Fig.\ref{zev}.
An evolution in the $T_{\rm d}$ as a function of redshift is has been observed by many studies using stacking analyses \citep[e.g.][]{Magdis2012,Magnelli2014,Bethermin2015,Ivison2016}, typically finding an evolution following $T_d\propto(1+z)^{0.32}$. Assuming $T_d=40\,{\rm K}$ at $z=3.2$, this evolution then predicts $T_d\sim\,43\,{\rm K}$ and $T_d\sim\,46\,{\rm K}$ at $z\sim4.3$ and $z\sim5.5$, respectively.

Relative to our fiducial value of $T_{\rm d}=40\,{\rm K}$, these assumptions would increase the IRX values of the observed samples by $\sim0.1\,{\rm dex}$, and $\sim0.2\,{\rm dex}$ at $z\sim4.3$ and $z\sim5.5$, respectively.
Clearly, a $T_{\rm d}$ evolution could thus partially compensate for the observed IRX deficit at $z>4$.
However, a much more extreme evolution of the dust temperature would be required to fully account for the median decrement in IRX values at $z>4$.
Nevertheless, it is clear that better constraints on the dust temperatures of high redshift galaxies will be crucial for future studies.

\section{Conclusions}

In this paper, we have presented an investigation of the IRX-$\beta_{\rm UV}$ relation of massive (\Mstar$\sim10^{10.7}$\Msun)
star-forming galaxies at $z\sim3.2$, through FIR continuum observations with ALMA.
We also explored a possible redshift evolution of the IRX--\uvbeta\ and the IRX--\Mstar\ relations by utilising public ALMA observations 
of star forming galaxies between $z\sim3$ and $z\sim6$. Our findings can be summarised as follows:

\begin{itemize}
\item The IRX--\uvbeta~relation of our sample is on average mostly consistent
	with that of local starburst galaxies derived by \citet{Meurer1999} and we can exclude an SMC-type extinction curve for the general star-forming $z\sim3$ galaxy (see Fig. 2, 3). However, as the stack of our ALMA non-detections shows, certain sub-samples of galaxies exist that are in better agreement with the SMC relation. 
\\
\item Our individually detected sources and our stacks span a large range in IRX values at fixed \uvbeta,  consistent with previous analyses at $z\sim3$. This indicates the presence of a very large dispersion around the average IRX--\uvbeta\ relation of up to $\pm0.5$dex. The same is observed for the  IRX--\Mstar\ relation (see Fig. 3,4).
\\	
\item With public ALMA observations of galaxies at $z\sim4$--6,
    we explored the redshift evolution of the IRX--\uvbeta\ and the IRX--\Mstar\ relations, which we find to be significant. Even though the current samples are still small, we find a strong, gradual decrease to lower IRX values at fixed \uvbeta\ or \Mstar\ with increasing redshift, with the $z\sim6$ sample generally lying an order of magnitude below our $z\sim3$ relations on average (see Fig. 4). 
\end{itemize}

The fact that our $z\sim3$ sample is consistent with the local M99 dust law, yet higher redshift samples appear to fall off this relation, is particularly interesting, as it could point to significant changes in the ISM and dust properties of high-redshift galaxies. It will be important to confirm the suggested evolution between $z\sim3$ and $z\sim6$ with future datasets.
In particular, our study demonstrates that individual IR detections of the dust emission from a statistical high-redshift sample are crucial due to considerable scatter around the 
average IRX--\uvbeta~that can not be fully captured by stacking analyses. Our analysis thus motivates a larger investment of ALMA time to study the dust emission of a statistical sample of normal high-redshift galaxies.

\section*{Acknowledgements}
The authors thank A. Ferrara, R. Bouwens, and N. Reddy for very helpful discussions related to this work.
This paper makes use of the following ALMA data:
 ADS/JAO.ALMA\#2013.1.00151.S, ADS/JAO.ALMA\#2013.1.00034.S, and ADS/JAO.ALMA\#2012.1.00523.S.
ALMA is a partnership of ESO
(representing its member states), NSF (USA) and NINS (Japan), together
with NRC (Canada) and NSC and ASIAA (Taiwan) and KASI (Republic of
Korea), in cooperation with the Republic of Chile. 
The Joint ALMA Observatory is operated by ESO, AUI/NRAO and NAOJ.
Based on data products from observations made with ESO Telescopes at the La Silla Paranal Observatory under ESO programme ID 179.A-2005 and on data products produced by TERAPIX and the Cambridge Astronomy Survey Unit on behalf of the UltraVISTA consortium. 
B.G. gratefully acknowledges the support of the Australian Research Council as the recipient of a Future Fellowship (FT140101202).
MTS acknowledges support from a Royal Society Leverhulme Trust Senior Research Fellowship (LT150041).
VS acknowledges support from the European Union's Seventh Frame-work program under grant agreement 337595 (ERC Starting Grant, 'CoSMass').
OLF acknowledges support from the European Research Council Advanced Grant ERC-2010-AdG-268107-EARLY.




\bibliographystyle{mnras}
\bibliography{./ref.bib}

%


\bsp	
\label{lastpage}
\end{document}